# `SEPhIA`: $< 1$ laser/neuron Spiking Electro-Photonic Integrated Multi-Tiled Architecture for Scalable Optical Neuromorphic Computing

Matěj Hejda[a,*], Aishwarya Natarajan[a], Chaerin Hong[a], Mehmet Berkay On[a], Sébastien d'Herbais de Thun[a], Raymond G. Beausoleil[a], Thomas Van Vaerenbergh[a]

[a]Large-scale Integrated Photonics (LSIP), HPE Labs, Hewlett Packard Enterprise

## Abstract

Research into optical spiking neural networks (SNNs) has primarily focused on spiking devices, networks of excitable lasers or numerical modelling of large architectures, often overlooking key constraints such as limited optical power, crosstalk and footprint. We introduce `SEPhIA`, a photonic-electronic, multi-tiled SNN architecture emphasizing implementation feasibility and realistic scaling. `SEPhIA` leverages microring resonator modulators (MRMs) and multi-wavelength sources to achieve effective sub-one-laser-per-spiking neuron efficiency. We validate `SEPhIA` at both device and architecture levels by time-domain co-simulating excitable CMOS-MRR coupled circuits and by devising a physics-aware, trainable optoelectronic SNN model, with both approaches utilizing experimentally derived device parameters. The multi-layer optoelectronic SNN achieves classification accuracies over 90% on a four-class spike-encoded dataset, closely comparable to software models. A design space study further quantifies how photonic device parameters impact SNN performance under constrained signal-to-noise conditions. `SEPhIA` offers a scalable, expressive, physically grounded solution for neuromorphic photonic computing, capable of addressing spike-encoded tasks.

---

[*]Corresponding author. Address: Hermeslaan 1A, Diegem, Belgium. Email: matej.hejda@hpe.com

# 1. Introduction

Neuromorphic computing [1] represents an emerging paradigm that relies on biologically-inspired methods and principles for computing and neuroscientific applications [2]. For computing, the objective is to realize more resource and energy-efficient computation, particularly for machine learning (ML), artificial intelligence (AI), and sensor data processing. In the hardware domain, neuromorphic chips utilize a varying set of neuro-inspired concepts, including near-memory and in-memory computing [3] as well as spike-based signalling and computing. The field of neuromorphic accelerators encompasses a broad variety of approaches from both commercial [4], [5], [6] and academic teams [7], [8], [9], [10]. Beyond electronics, photonic computing is a rapidly developing field, focusing on the development of workload-specific accelerators that seize some of the highly desirable properties of photonics [11], primarily for AI acceleration [12]. Photonic neuromorphic and spike-based computing [13] is comparably less mature, yet undergoing notable growth in research interest thanks to its promise of computing with ultra-low power [14], extensive bandwidth [11] surpassing that of electronics, and noise-robustness of spiking neural networks (SNNs) [15]. Besides all-optical approaches, optoelectronics allow us to seize the advantages from both of the signal processing domains [16]: the high degree of parallelism and (nearly) lossless communication offered by the optical components, with the maturity, robustness, and readily accessible nonlinearities in the electronic domain. Currently, the arguably most explored aspect of spike-based neuromorphic photonics are the neurons (i.e., exploration of excitable and spiking dynamics in photonic and optoelectronic devices), with comparably fewer works at circuits at architecture level. Particularly in contrast to general optical computing, comprehensive architecture-focused studies of integrated photonic SNNs are currently less explored.

To unlock the full practical potential of photonic spike-based computing, there is a need for neuromorphic photonic architectures that are (a) scalable under realistic consideration for (current) photonic technology, and (b) validated with comprehensive, true-to-hardware, end-to-end SNN models. In terms of scalability, a hardware-software co-designed modelling framework for optical SNNs has been recently reported using spiking DFB-SA lasers and Mach-Zehnder modulator (MZM) cells [17], demonstrating multiple types of functional photonic cells for realizing spiking convolutional neural networks (CNNs). While certain typical aspects of analog computing are considered (such as limited bit precision of MZMs), other proposed aspects, such as optical power limitations regarding the use of massive optical fan-out (100+ devices), remain a significant challenge. In terms of neuromorphic photonic architecture validation and benchmarking, various simplified approaches are often used. These include embedding of a smaller functional photonic (or photonic-like) block(s) in a much larger and complex ML model or pipeline ([18], [19]), or making an existing ML model 'photonic' solely by modifying some of its parameters based on photonic devices [20]. However, in the first case, the reported figures of merit (such as classification accuracies) are often heavily determined by the digital ML model rather than the photonic blocks. Meanwhile in the latter case, sole reliance on parameters and characteristics in conventional ML models does not directly capture many of the challenging aspects of optical computing, including limited precision, noise, photonic components' physics, limited (optical) power budgets, and scalability constraints.

To address both of the key points above, we propose `SEPhIA` (Spiking Electronic-Photonic Integrated Architecture). `SEPhIA` is a microring modulator (MRM)-based, non-coherent, wavelength division multiplexing (WDM)-enabled hybrid photonic-electronic neuromorphic architecture that focuses on scalability and practical feasibility, and addresses a multitude of challenges observed in previously proposed photonic neuromorphic devices and architectures, such as:

- footprint and scalability limitations of the most common photonic neurons (which typically assign a dedicated laser per each individual neuron) $\rightarrow$ *by using a shared multi-wavelength laser with a single MRM per neuronal unit, the footprint per neuron is significantly reduced (see Section 3 for details);*
- challenges related to all-optical approaches, including limited fan-in of coherent devices, or lack of spike inhibitory functionality $\rightarrow$ *optoelectronic processing with balanced photodetection typically alleviates both of these challenges;*

- in some cases, limited dynamical expressivity of spiking dynamics in nonlinear physical devices such as excitable lasers → *analog CMOS neuron dynamics can be controlled and tuned conveniently by tuning of circuit parameters [21];*
- architecture scaling limitations of analog optical computing → *we carefully consider optical power budget and frequency domain limitations of current photonic technology, and devise our `SEPhIA` architecture in a multi-tiled, sparsely connected architecture that accounts for these physical constraints while maintaining good performance (see Section 3 for details).*
- energy overheads from use of high-resolution domain converters (such as analog-to-digital converters, ADCs), which can represent a significant part of photonic accelerator energy budget [22] → *spike-based optoelectronic computing allows for all-analog (ADC-less) signal processing within the architecture;*

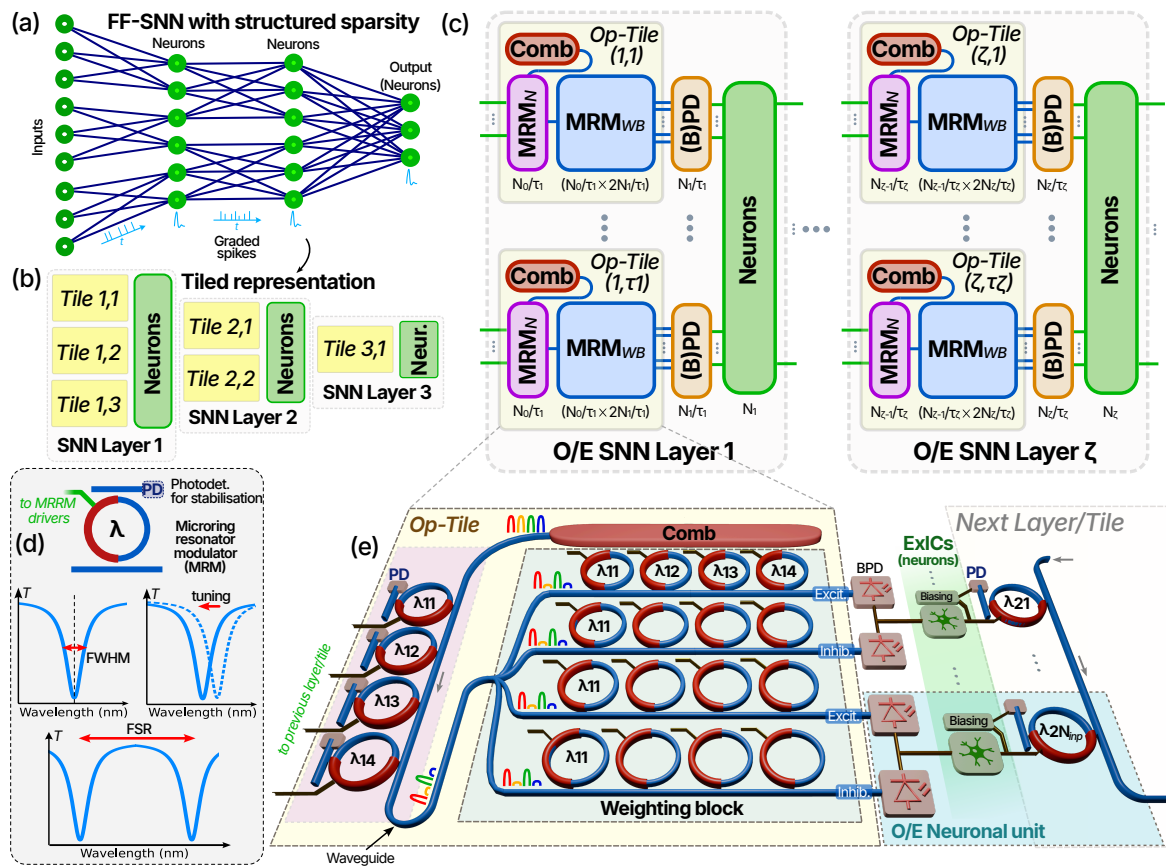

Figure 1: (a) Simplified diagram of a structurally sparse feed-forward SNN. (b) Multi-tiled SNN architecture implementing the sparse SNN. (c) A schematic diagram of an example of `SEPhIA`, the proposed multi-tiled WDM-enabled O/E-SNN hardware architecture combining photonic weighting circuit with analog electronic excitable neurons. (d) Key characteristics of microring resonator modulators (MRMs), a key building block of the `SEPhIA` architecture. (e) A more detailed schematic at the neural network layer (and Op-Tile) level, highlighting the two key functional blocks: (1) the weighting block, here depicted as a $4 \times 4$ all-pass MRMs weight bank with broadcasting, and (2) the opto-electronic neuronal unit, consisting of a shared multi-wavelength source (comb laser) per neural layer, BPDs and analog excitable CMOS circuits coupled to array of MRMs for wavelength-selective E/O conversion of spikes.

The `SEPhIA` analog neuromorphic photonic-electronic architecture implements all the required functional blocks for realizing a deep SNNs with weighting using graded spikes [23]. Due to the lossy nature of realistic photonic components and our design choice of not implementing on-chip optical amplifiers within the architecture (primarily due to their footprint), there's an inherent limitation on the photonic circuit size (see further analysis in Section 3.1). Acknowledging this, we realize a multi-tiled opto-electronic (O/E) SNN which aims to represent SNNs with block-diagonal sparse weighting matrices (see example in Figure 1(a)). Sparsely connected deep neural networks offer significantly better suitability for integration in hardware while often incurring minimal use-case dependent decrease in classification accuracy [24].

These block-diagonal components of a given weighting matrix within an SNN layer can be considered as individual, parallel functional circuits called *tiles*. The same sparse SNN in the form of tiles is shown in Figure 1(b). In our architecture, these tiles will be referred to as Op-Tiles (optical tiles). Unlike the case of time-domain multiplexed tiled optical processing of matrices, [25], our tiles represent a case of hardware parallelism. The multi-tiled deep O/E-SNN architecture is shown in Figure 1(c), with more details of a single Op-Tile shown in Figure 1(e). Each Op-Tile contains **(a)** a single, multi-wavelength optical source (such as a comb laser), **(b)** a set of MRMs, shown in Figure 1(d)), followed by **(c)** MRM-based, WDM-enabled integrated photonic signal routing and weighting circuit (an MRM weight bank [26], previously experimentally demonstrated as suitable for optical spike weighting [27]). Following that, the weighted optical signals are summed up on **(d)** pairwise balanced photodetectors (BPDs), whose output is considered as the output of the Op-Tile. Each Op-Tile has two defining size parameters: number of inputs $N_{\text{inp}}$ and number of outputs $N_{\text{out}}$. $N_{\text{inp}}$ is defined by the number of available WDM channels, while $N_{\text{out}}$ is flexible and can be adjusted. In the simplest case where $N_{\text{inp}} = N_{\text{out}}$, we can refer to the Op-Tile size $N_T \equiv N_{\text{inp}} = N_{\text{out}}$. All the outputs from a given $l$-th layer of $\tau_l$ parallel Op-Tiles drive a set of analog, electronic excitable integrated CMOS circuits (ExICs, $n_l^{\text{ExIC}} = \sum_{k=1}^{\tau_l} N_{\text{out}_k}$), which provide the neuron-like excitable dynamics (=spiking). This set of $\tau_l$ parallel Op-Tiles and the corresponding $n_l^{\text{ExIC}}$ electronic neurons constitutes a single layer of the feed-forward SNN, with total of $\zeta$ layers.

## 2. Results

To validate the idea of our proposed `SEPhIA` neuromorphic optoelectronic architecture, we provide a comprehensive set of different types of numerical simulations. First, we utilize a Verilog-A based electronic-photonic co-simulation model for time-domain modelling of the O/E neuronal units (CMOS+MRM), including a compact model of a MOSCAP-based silicon MRM that captures its electrical and optical dynamics. Second, we implement a multi-layer feed-forward O/E-SNN by leveraging the combination of `snnTorch` [28] framework and our custom `pyTorch`-based frequency-domain functional simulator of photonic neural networks based on compact photonic device models. Thanks to the use of `pyTorch`, our O/E-SNN model is end-to-end differentiable and can utilize the `autograd` engine for true-to-hardware gradient-descent-based O/E-SNN model training.

### 2.1. Device-level co-simulation of the O/E neuronal unit

First, we focus on the O/E neuronal unit as shown in Figure 1(e). There are two main functional parts in an O/E neuronal unit of `SEPhIA`: a photodetector-coupled analog ExIC, and an MRM. The ExIC implements the nonlinear spiking functionality and generates the action potential (spikes). We utilize an adaptive exponential leaky integrate-and-fire (AdEx LIF) circuit model (Figure 2(a), [21], [29]), with optical inputs enabled via a pair of balanced photodetectors (BPDs) for both excitatory (+) and inhibitory (-) functionality. The ExIC exhibits neural heterogeneity through tuneable parameters, enabling the O/E spiking neurons to be configured such that they can produce various types of responses. In this demonstration, a set of $n = 4$ individual ExICs is coupled to a set of $n = 4$ individual MRMs on a shared bus waveguide (Figure 2(b)). In the current architecture, neural outputs are encoded at the through ports of the MRMs (Figure 2(c)). Various methods of photonic-electronic co-integration are depicted in Figure 2(d).

The transient responses of the four ExICs simulated in Verilog-A when subject to current inputs from Figure 2(e-g) are shown in Figure 2(h-j). We demonstrate three different kinds of spiking behaviors [30] at 1GSpike/s rates: in the first regime (Figure 2(e,h,k)), the circuit produces a tonic (repeated, regularly spaced) spiking under a constant input with weak adaptation, representing neurons that sustain continuous firing at nearly constant rates, similar to a conventional LIF neuron. In the second case (Figure 2(f,i,l)), a spike-frequency adaptation is observed as the neuron is stimulated with a pulsed input current. With each spike, the spike-triggered adaptation mechanism kicks in, increasing the inter-spike interval, and gradually reducing the firing rate as the neuron adapts to the input pulse. Third (Figure 2(g,j,m)), we demonstrate bursting dynamics, with a slower adaptation, in which clusters of continuous spikes are followed by quiescent intervals. For each of the regimes of the ExIC, we provide the corresponding time traces (Figure 2(k-m)) of photonic-electronic Verilog-A co-simulation for the coupled MRM-ExIC system in the neuronal units. These traces show the optical power per wavelength at the output of the shared bus waveguide for the case where the $n = 4$ ExIC circuit outputs simultaneously directly modulate the MRMs, with a model case of $P_\lambda = 1$ mW, and demonstrate the viability of ExIC-driven MRMs for WDM spiking photonic architectures.

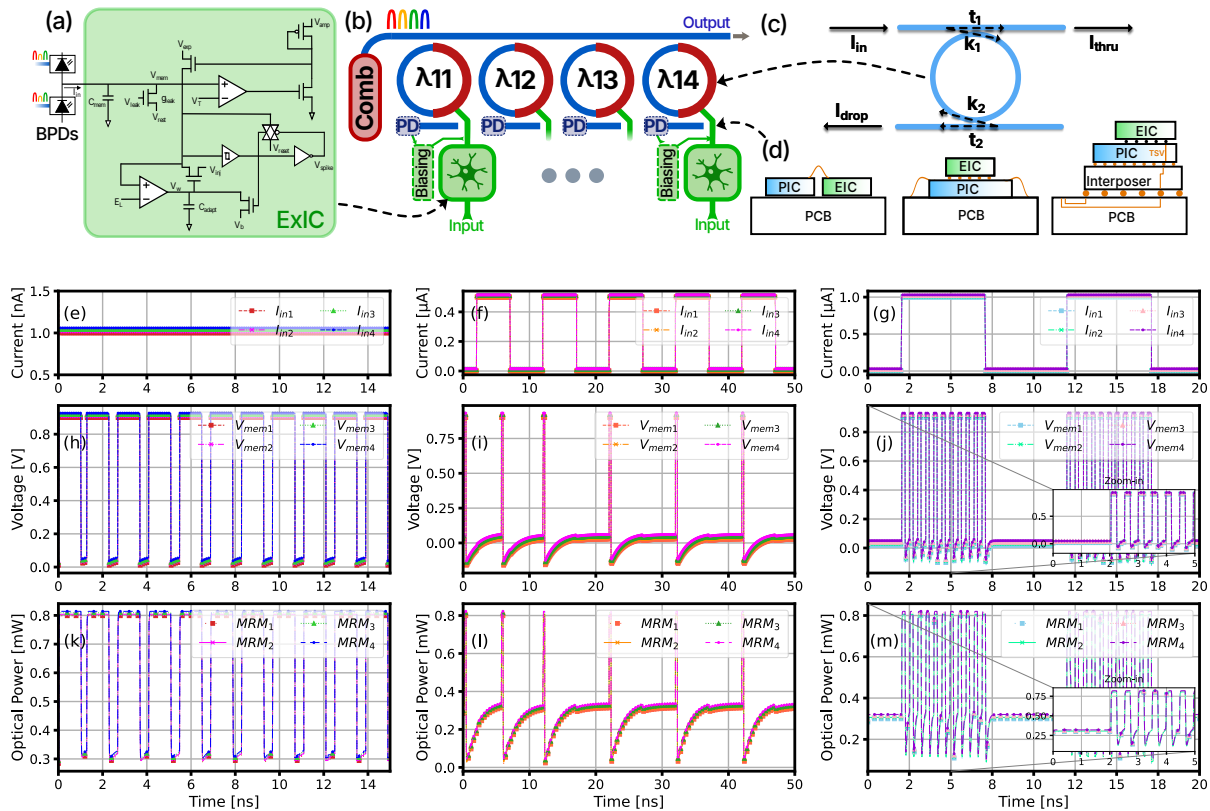

Figure 2: Schematics and operation of O/E neuronal unit: (a) Circuit schematic of the neuron model in the ExIC. (b) A simplified schematic of four parallel O/E neuronal units on a single shared bus waveguide. (c) Schematic of an add-drop microring resonator. (d) Various possible solutions for optical-electronic coupling and co-integration. (e-m) Time traces of Verilog-A simulations: (e-g) The used input signal (current) to the O/E neuronal unit; (h-j) Voltage responses from the excitable analog CMOS (ExIC) circuits. (k-m) Neural MRM through-port output (optical power of a single WDM channel) from the co-simulation of full O/E neuronal units.

Figure 3 numerically demonstrates the operation of these MRMs, using a numerically generated comb laser spectrum (in green, see Supplementary info), an ideal transmission spectrum of an MRM (dashed lines) and a pass-through spectrum of multiple MRMs (dark blue). Due to the notch frequency filtering nature of the MRMs, some degree of crosstalk among adjacent modulation channels is inevitable. To evaluate the optimal amount of modulation-induced resonance shifting in the MRM with respect to crosstalk, we aim to maximize a figure of merit represented as a sum of extinction ratios (ER) of two adjacent WDM channels, modulated by two independent MRMs. The figure of merit is shown in Figure 3(e) as a function of the ring $Q$-factor and the on-off keying MRM resonance shift $\Delta\lambda_{\text{reso}}$, applied to two MRMs operating on adjacent WDM channels. For the depicted case of $Q = 7.5$K, $\Delta\omega = 63$ GHz at 1310 nm, the frequency domain crosstalk-optimized shifting range appears to be $\Delta\lambda_{\text{optimal}} \approx 210$ pm. This value (and correspondingly scaled values for $\Delta\omega = 50$ GHz and $\Delta\omega = 100$ GHz) guides the MRM resonance shifting range choices $\Delta\lambda_{\text{optimal}}$, $\Delta\lambda_{\text{max}}$ (for weighting) in Section 2.2.

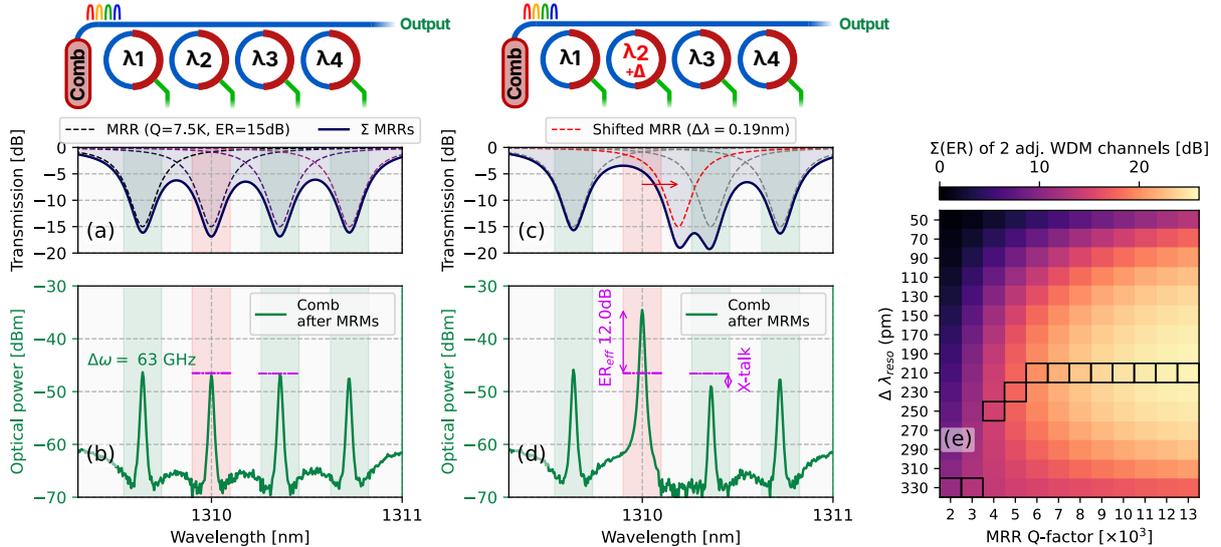

Figure 3: Demonstration of frequency domain crosstalk as observed in the MRM-based O/E neuronal units, in a model case of $n = 4$ individual neural MRMs. (a,b) A steady state of the system, where no neuronal unit is excited. (c,d) A state where the second O/E neuron is excited, and the MRM is briefly switched, demonstrating both the effective ER, as well as the crosstalk. (e) A figure of merit (FoM) for finding the optimal resonance shift. The FoM, which we aim to maximize, is a sum of the total optical powers on two adjacent WDM channels, where the MRM corresponding to the WDM first channel is modulated towards the second channel. For $\Delta\omega = 63$ GHz, we observe max(FoM) (denoted as black squares) for $\Delta\lambda = 210$ pm for all ring $Q$-factors above 6K.

## 2.2. Hardware-aware trainable O/E-SNN model

The full O/E-SNN training procedure flow diagram is shown in Figure 4(a). Thanks to the end-to-end trainable nature of the developed model, the training procedure directly adapts to key aspects of the photonic hardware, such as non-uniform WDM optical channel powers from the multi-wavelength source, the actual optical powers available for each WDM channel, the specific transfer function of the MRMs, the frequency-domain crosstalk across MRMs, and the optical power attenuation at each photonic component. The O/E-SNN is trained using supervised training with backpropagation through time (BPTT) [31] over `n_timesteps` = 35 discrete timesteps using a surrogate `arctan` function for gradient calculation of LIF neurons during the backward pass [32]. Example traces recorded during a set of training procedures are shown in Figure 4(b,c), with O/E-SNN training losses and validation accuracy recorded over 15 epochs for $n = 5$ repeated training runs. In this case, we assume all the MRMs' parameters as $Q = 10K$ and ER = 15 dB, and WDM channel spacing of $\Delta\omega = 100$ GHz. Full set of simulation parameters is available in the Supplementary information. For the classification problem, we are using a subset (4 classes, similarly to [33]) of the dimensionally-reduced fashion image dataset [34] encoded into rate-coded spike trains (see Supplement for details). Both the curves confirm that the true-to-hardware O/E-SNN model is well trainable, reaching (and exceeding) 90% classification accuracy on the selected problem.

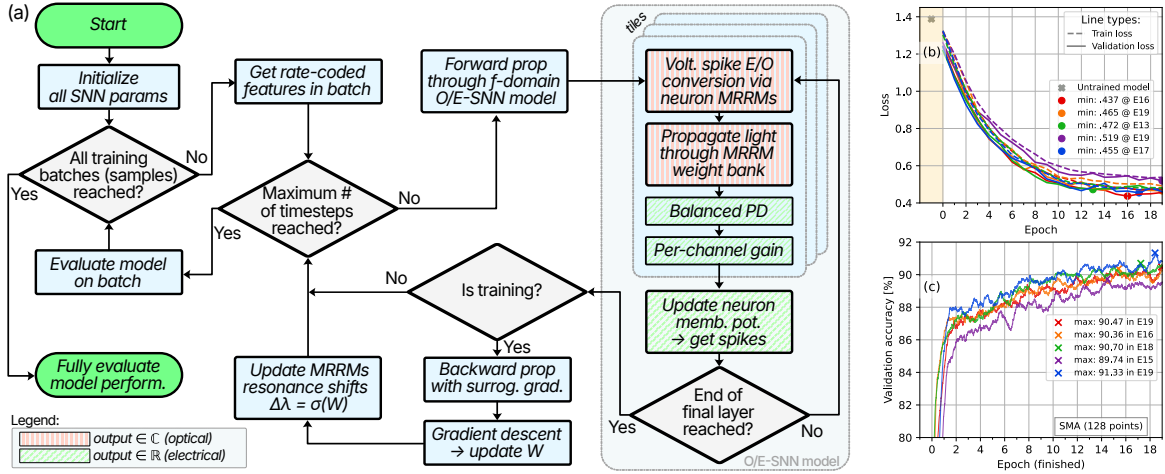

Figure 4: Training of the O/E-SNN. (a) Flow chart of the O/E-SNN model training procedure. (b) Training and validation losses (logged per epoch) during training. (c) Validation accuracy (logged per batch) during training. Plotted as a simple moving average (SMA) over 128 points.

In particular, the O/E-SNN trained in this example is a two-layer O/E-SNN, where the first layer contains two parallel Op-Tiles (as shown in Figure 1(e)), each of size $N_T = 16$ (corresponding to a $16 \times 16$ all-pass MRM weight bank), $n_{l=1}^{\text{ExIC}} = 16$ spiking neurons, and the second layer contains a single Op-Tile with $N_{\text{inp}} = 16$, $N_{\text{out}} = 8$ (corresponding to a $16 \times 8$ weight bank) followed by $n_{l=2}^{\text{ExIC}} = 4$ spiking neurons. The full tiled architecture with all the main functional blocks is also shown in the schematic in Figure 5(a). As an optical source for each Op-Tile, we assume an idealized frequency comb laser source with $n_\lambda = 16$ WDM channels, with uniformly randomly distributed peak optical power within a 2dB band from the maximum optical power $P_\lambda$ for each comb line. Figure 5(b) depicts the internal state variables (membrane potentials) of all the LIF neurons (ExICs) in the first layer (for both Op-Tiles in the layer, divided by thick black line in the plot). Figure 5(c) depicts the corresponding spikes at the output of all the O/E neuronal units between first and second O/E-SNN layer. The slight variation in observed output optical powers for different neurons in Figure 5(c) comes from the non-uniform optical power distribution of the WDM channels at the multi-wavelength source, and is adapted for, by the physics-aware training procedure. Figure 5(d) depicts the membrane potentials at the final layer of neurons, and Figure 5(e) depicts the classifier prediction (based on cumulative count of output spikes per each neuron) over all the timesteps. We can observe the cumulative counts reaching maximum of $n = 12$ spikes for neuron 1, corresponding to a correct classification of the input sample. Figure 5(f) then depicts the confusion matrix for the trained O/E-SNN on the test-set, corresponding to total classification accuracy of 91.35%. We can compare this to two feed-forward SNN baselines of the same size: a fully-connected two-layer (FC-SNN), and a SNN with structured sparsity (Table 1). Using the same training procedure and same dataset, we achieve 93.65% classification accuracy for the FC-SNN, and 92.97% classification accuracy for the structurally-sparse SNN. This demonstrates that introducing a degree of structured sparsity into the model for the given task yields a classification accuracy that's comparable to a fully-connected model, and also confirms that the performance of our physics-aware O/E-SNN model with noise and crosstalk effects is comparable to that of an ideal SNN, only trailing behind the conventional SNN by $\approx$ 1 percentage point.

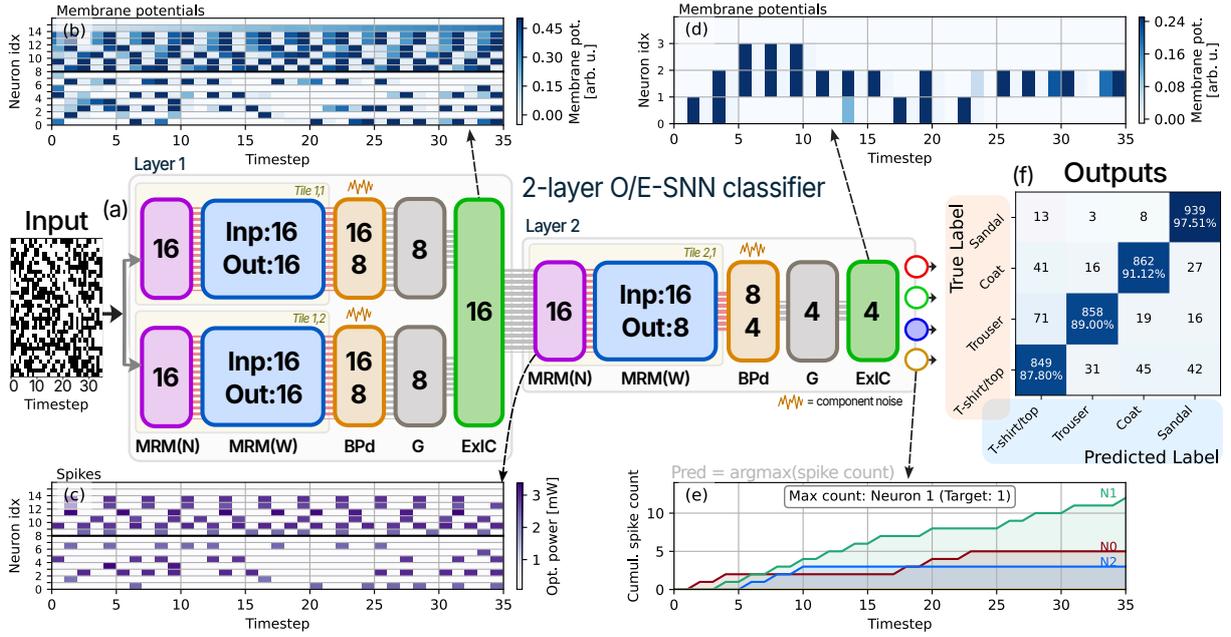

Figure 5: Schematic diagram of the two-layer O/E-SNN architecture. (a) A simplified diagram showing the sizes for various parts of the O/E-SNN. (b) Membrane potentials at the LIF neurons in the first layer. (c) Output optical spikes (each at a separate WDM channel) after the neural block between the first and second layer. (d) Membrane potentials at the LIF neurons in the second layer. (e) Cumulative spike counts at the O/E-SNN output. (f) Confusion matrix of the O/E-SNN classifier (total classification accuracy is 91.35%).

Furthermore, the O/E circuit performance is influenced by a complex interplay of various physical characteristics of the system, including MRM parameters ($Q$-factor, extinction ratio, insertion loss), PD parameters (thermal noise, shot noise and noise equivalent power) and multi-wavelength source characteristics (WDM channel spacing $\Delta\lambda$, optical power levels, mode linewidths, noise), among others. The presented model allows us to directly explore the design space of these parameters and their interplay on the selected figure of merit (classifier accuracy) of the O/E-SNN. For WDM channel spacing $\Delta\omega$, we explore three different options: Ⓐ $\Delta\omega = 100$ GHz; Ⓑ $\Delta\omega = 63$ GHz; and Ⓒ $\Delta\omega = 50$ GHz. Corresponding MRM shifting ranges are set for each $\Delta\omega$. For each $\Delta\omega$, we also explore multiple model sets of MRM design parameters with various $Q$-factors and extinction ratios (ER): higher $Q$ MRMs including ① $Q = 10K$ and ER = 15 dB, motivated by realistic MOSCAP MRM device parameters [35]; ② $Q = 10K$ and ER = 6 dB; and lower $Q$-factor MRM as ③ $Q = 5K$ and ER = 6 dB. In total, this yields 9 sets of design parameters. In all cases, we assume a fixed set of ideal PD parameters, which yields noise equivalent power (NEP) of $3.6 \times 10^{-11} \text{W}/\sqrt{\text{Hz}}$, or $\approx -27$ dBm minimum detectable power at 2.5 GHz bandwidth under dark conditions (see Supplementary info for more details).

We can observe the mean O/E-SNN validation accuracies during training for all the parameter sets in Figure 6. A total of $n = 5$ independent repeated training runs have been performed for each set of parameters. We can see that the lowest overall validation loss was achieved for parameter set Ⓐ① ($\Delta\omega = 100$ GHz, $Q = 10K$, ER = 15 dB), that is, for the broadest WDM spacing, and for MRMs with high-$Q$ and high ER. In such case, the frequency domain crosstalk effects have the lowest influence. An overall trend of increase in validation losses can be observed as $\Delta\omega$ is decreased, indicating that tighter channel spacing brings crosstalk-related impairments to the system performance. Interestingly, it can be observed that for the lowest tested $\Delta\omega = 50$ GHz in Figure 6(c), better performance (lower validation loss) is achieved with ② (ER = 6 dB) rather than ① (ER = 15 dB), indicating that for ultra-dense WDM channel spacing, lower MRM extinction ratio can yield a better performing system, which was not the case for larger WDM channel spacing. This further highlights the complexity of interplay of various physical effects at the device and circuit level, which have to be considered when realizing photonic computing architectures. The minimal observed validation losses are summarized in Table 1.

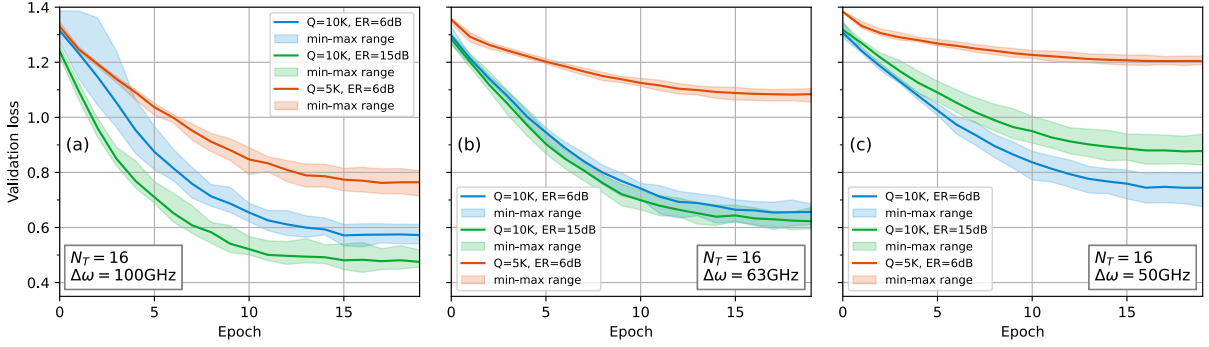

Figure 6: Mean validation loss traces during training of the multi-tiled, two-layer O/E-SNN with various WDM channel spacing and MRM parameters.

Finally, we explore the robustness of the trained O/E-SNN against changes in WDM channel optical power. Figure 7 shows the sweep results for all 9 parameters sets discussed previously. In all the cases, the network was trained with a peak power per WDM channel $P_\lambda = 6$ dBm (denoted as dash-dotted red vertical line in the figure, see Section 3.1). After the training, a test set classification was performed while sweeping $P_\lambda$ from 12 dBm to $-4$ dBm, effectively bringing the system outside of operational conditions it was trained for. In the majority of the cases, we observe that highest accuracies are achieved for powers slightly above the selected training $P_\lambda$, which means the selected $P_\lambda$ represents the lower bound of optical powers under which the system can perform well. In agreement with previous data, we observe the highest degree of robustness for Ⓐ ① ($\Delta\omega = 100$ GHz, $Q = 10K$, ER $= 15$ dB). Furthermore, we observe that with decreasing WDM channel spacing, the performance of ② gradually becomes comparable to ①, which is in agreement with the validation loss data in Figure 6, and we also observe that in the case of Ⓒ ③ (Figure 7(c)), the O/E-SNN performance is effectively capped by the physical effects (cross-talk, noise), further emphasizing that practical challenges can arise for ultra-dense WDM optical computing systems. Interestingly, we also observe slight accuracy drop at optical powers exceeding the value used for training. This can be attributed to the presence of nonlinearity at the photodetectors (where photodetector current $I_{\text{out}} \propto P \propto E_{\text{field}}^2$). The maximal observed training accuracies for each parameter set, as well as the accuracy change when $P_\lambda$ is decreased by 3 dB from the optical power corresponding to the maximum accuracy (as a simple robustness measure) are shown in Table 1.

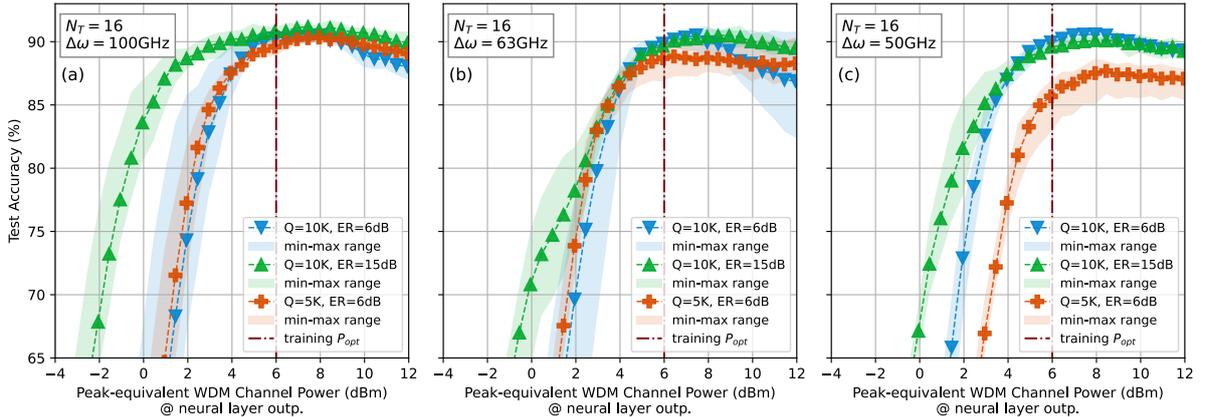

Figure 7: Dependence of O/E-SNN accuracy as a function of the peak optical power of a single WDM channel $P_\lambda$ at the Op-Tile multi-wavelength source stage.

Table 1: Performance metrics of the two-layer O/E-SNN during the design space exploration.

| Model | $\Delta\omega$ | MRR_Q , MRR_ER | Test Acc. (best) @ trained $P_{\text{opt}}$ | $\Delta$(Test Acc.) (best) @ $P_{\text{opt}} - 3$ dB | min val. loss |
|---|---|---|---|---|---|
| baseline SNN (FC) | | | (93.65%) | - | - |
| baseline SNN (sparse) | | | (92.97%) | - | - |
| O/E-SNN (sparse) | Ⓐ 100 GHz, | ① 10K, 15dB | 91.64% | $-0.62$%pt | 0.437 |
| | Ⓐ 100 GHz, | ② 10K, 6dB | 90.86% | $-2.89$%pt | 0.530 |
| | Ⓐ 100 GHz, | ③ 5K, 6dB | 90.55% | $-3.44$%pt | 0.714 |
| | Ⓑ 63 GHz, | ① 10K, 15dB | 90.36% | $-3.72$%pt | 0.592 |
| | Ⓑ 63 GHz, | ② 10K, 6dB | 91.04% | $-4.22$%pt | 0.637 |
| | Ⓑ 63 GHz, | ③ 5K, 6dB | 89.64% | $-3.29$%pt | 1.073 |
| | Ⓒ 50 GHz, | ① 10K, 15dB | 90.10% | $-3.25$%pt | 0.826 |
| | Ⓒ 50 GHz, | ② 10K, 6dB | 90.94% | $-3.01$%pt | 0.695 |
| | Ⓒ 50 GHz, | ③ 5K, 6dB | 87.89% | $-12.79$%pt | 1.188 |

## 3. Discussion

A first general aspect to consider in terms of an integrated photonic spiking architecture is the area footprint $S$. Arguably the most commonly explored case of spiking photonic laser neurons are multi-section DFB [17] or Fabry-Perot (FP) lasers [36], [37]. In the case of FP lasers, both referenced works report approximately 1.5 mm cavity length. Since a single device with footprint $S_{\text{laser}}$ is used per neuron, the footprint of neural layer with $N$ neurons scales $S \propto (N \cdot S_{\text{laser}})$. In the case of SEPhIA, the cavity lengths of the used multi-wavelength sources are comparable (0.4 mm to 1.6 mm in [38], 1.4 mm [39]), but the footprint of neurons within SEPhIA scales only with the number of MRMs with footprint $S_{\text{MRM}}$ as $S \propto (S_{\text{laser}} + N \cdot S_{\text{MRM}})$. Given that in silicon photonics, MRM diameters can be $< 50$ um, and therefore $S_{\text{MRM}} \ll S_{\text{laser}}$, our shared WDM-laser approach offers significantly reduced footprint and improved scalability prospects. Alternatively, an array of integrated microring lasers [40] could be considered as a footprint-viable optical source option, where the current primary limiting factors are the low maximal achievable levels of output optical power.

Aside from the footprint and photonic technology induced limitations, there are two additional primary constraining aspects related to the scaling of individual Op-Tiles: (i) the optical power loss budget limitations and the (ii) frequency domain limitations.

### 3.1. Op-Tile scaling estimation: optical power budgets

For a single Op-Tile of size ($N_{\text{inp}}$, $N_{\text{out}}$), we can consider that a single WDM channel signal from the shared multi-wavelength source must first pass through a series of $N_{\text{inp}}$ MRMs that encode the spikes from a previous layer, then the signal is optically (spatially) broadcast into $N_{\text{out}}$ waveguides and finally passes through $N_{\text{inp}}$ weighting MRMs before detection at the (B)PDs (Figure 8). Therefore, if we assume a given insertion loss (MRR_IL) for all the MRMs in the architecture and if we assume the ideal case of lossless optical power broadcasting, we can consider the total minimal power reduction (without spiking or weighting) in dB as $2N_{\text{inp}} \cdot \text{IL}_{\text{MRR}} + 10\log(N_{\text{out}})$. This is plotted in Figure 8(c) for $N_{\text{in}} = N_{\text{out}} \equiv N_{\text{T}}$ for various values of MRR_IL. This represents the power budget solely from the power splitting and losses at each component along the signal pathway, and does not represent additional optical power requirements due to channel crosstalk or similar effects.

To get an estimate for optical power per WDM channel, we can take a recent O-band quantum-dot frequency comb laser study, which reported 2.2 THz bandwidth laser with channel spacing $\Delta\omega$ between 25-100 GHz channel spacing, with optical power of 3.5 mW/mode for 50 GHz spacing [38]. Therefore, to remain within the same magnitude of optical powers, we highlight $P_\lambda = 6$ dBm/mode ($P_\lambda \approx 3.98$ mW/ mode) in Figure 8(d) as an expected upper limit for the optical power source within state-of-the-art without any additional optical amplification. The same optical power value was also previously used in other MRR-based integrated optical computing works [18]. We also assume a requirement of a dynamic range at the PD stage that enables 4 bits of power resolution [41]. For example, if we assume the previously discussed low 0.2 dB insertion loss at each MRM (orange line in Figure 8(d)), we see that the Op-Tile size is optical power-limited at approximately $N_\text{T}^\text{P-lim} \leqq 16$. This estimation currently does not consider waveguide propagation losses.

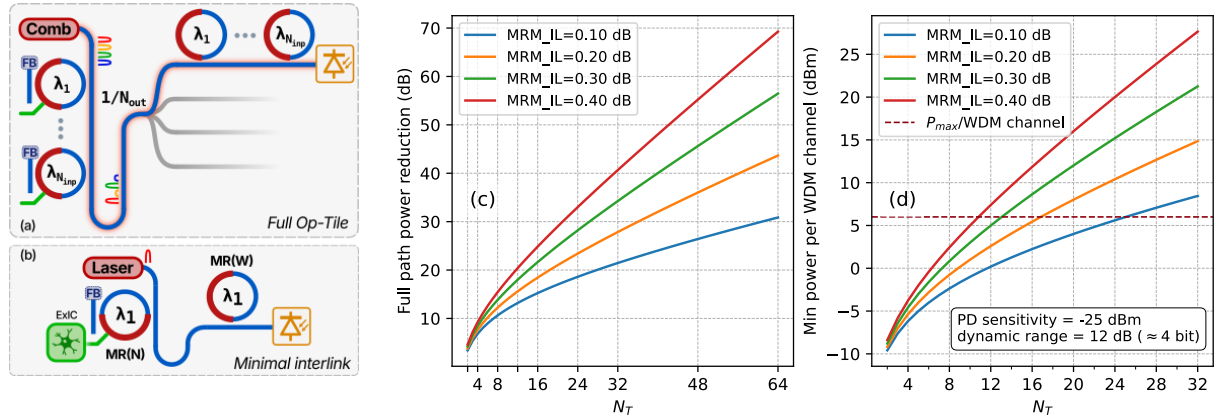

Figure 8: (a) Schematic diagram of a selected optical signal path through a single Op-Tile of size ($N_\text{inp}$, $N_\text{out}$). (b) Minimal O/E spiking interlink example. (c) Relative optical power decrease for a signal traversing the Op-Tile, plotted for various values of MRM IL. (d) Expected required minimal peak optical WDM channel power $P_\lambda$ for a single optical signal (at a single wavelength) traversing the Op-Tile.

### 3.2. Op-Tile scaling estimation: frequency domain constraints

In the frequency domain: to achieve independent control over all WDM channels, we are constrained by a single free spectral range (FSR) of used MRMs. We can consider a model case of a single SOI MRM [35] with radius $R = 10\mu$m and $\text{FSR}_\text{MR} = 7.518$ nm = 1.306 THz at 1310 nm. Using previously reported characteristics of heterogeneous quantum dot (QD) comb lasers [35], we can consider line spacings $\Delta\omega = 63/31.5/15.5$ GHz, and more recently $\Delta\omega = 100/50/25$ GHz [38]. Using a simple formula to get the maximum number of resonances within a single MRM FSR range $N_\text{f-lim} = \frac{\text{FSR}_\text{MR}}{\Delta\omega}$, we obtain numbers ranging from $N_\text{f-lim} = \frac{1306}{100} \approx 13$ to $N_\text{f-lim} = \frac{1306}{15.5} \approx 84$. While tighter packing of frequency channels towards ultra-DWDM (Dense WDM) approaches [42] increase the achievable tile size, they also introduce significant challenges related to channel crosstalk, requirement for high-$Q$ MRMs, and limited power per channel. Limitations related to FSR can also be addressed with the use of interleavers [43], or through exploiting multiple FSRs [44].

By considering the Op-Tile scaling limitation estimates based on state-of-the-art MRMs and multi-wavelength sources from the perspective of necessary optical power per each WDM channel ($N_\text{T}^\text{P-lim} \leqq 16$) and from the perspective of FSR-limit in the frequency domain (the most conservative $N_\text{f-lim} \leqq 13$ for the broadest WDM spacing $\Delta\omega = 100$ GHz without the use of interleavers or multiple FSRs), we believe the tile size $N_\text{T} = 16$ selected in our simulations represents a realistic, representative example.

### 3.3. Power-consumption analysis

One of the most frequently used figures-of-merit in neuromorphic photonics is the energy-per-spiking event. At the neuronal unit level, there are power consumption contributions from both the electronic ($P_E$) and optical ($P_\lambda$) domains. To enable accurate estimation of energy consumption $P_E$ in implementations of adaptive LIF neurons, we have developed a power-aware model that explicitly accounts for all major current components contributing to both dynamic and static power dissipation. The membrane capacitance current $I_\text{cap}$ models the charging/discharging of the membrane node and dominates the dynamic power during spike initiation and repolarization, calculated as, $I_\text{cap} = C_\text{mem} \cdot \frac{dV_\text{mem}}{dt}$. The leakage current $I_\text{leak}$ models the steady-state conductance pulling the membrane potential toward its resting potential, given by $I_{\text{leak}} = g_{\text{leak}}(V_{\text{mem}} - E_{\text{L}})$, and contributes to static power even in the absence of spiking activities. The exponential spike current $I_{\text{spike}}$ captures the activation of channel dynamics near threshold, defined as $I_{\text{spike}} = g_{\text{leak}} \cdot \Delta_{\text{T}} \exp\left(\frac{V_{\text{mem}} - V_{\text{T}}}{\Delta_{\text{T}}}\right)$, and dominates the dynamic power during spike onset. To reflect adaptation dynamics, the model also includes an adaptation capacitance $C_{\text{adapt}}$ that captures the transient charging/discharging during spikes, $I_{\text{adapt}} = C_{\text{adapt}} \cdot \frac{dV_w}{dt}$, where $V_w$ is the adaptation-state potential. This current modeling the continuous adaptation dynamics contributes an additional component to the dynamic power at spike events. In addition, the model includes the buffer-driving current for the microring's MOSCAP load, $I_{\text{buf}} = C_{\text{MRM}} \cdot \frac{dV_{\text{spike}}}{dt}$, which accounts for the extra energy required to drive the spike output into the optical modulator interface. All these currents are explicitly sourced from the supply rail to ensure realistic power accounting during SPICE-level simulations. The model also separates dynamic and static power contributions and integrates the total instantaneous power over each spike event to estimate the energy-per-spike, which is directly observable at the model output. Based on transient simulations, the proposed electronic neuron circuit (including 10 fF of $C_{\text{MRM}}$) consumes an average power of approximately $P_E = 4.586$ uW under typical operating conditions. In terms of optical power, our energy/spike metric has a degree of flexibility (by assuming adjustable $P_\lambda$ from the multi-wavelength source), and is considered from the perspective of a functional architecture. From Figure 7(a), we can assume $P_\lambda = 4$ dBm/channel ($\approx 2.5$ mW/channel in continuous-wave (CW) operation; this is the optical WDM channel power where full O/E-SNN accuracy is still maintained at $\approx 90\%$ before the accuracy roll-off, for the O/E-SNN with $\Delta\omega = 100$ GHz and MRM parameters $Q = 10K$ and ER = 15 dB).

The total required power for operating the O/E neuronal unit within the complete O/E-SNN with Op-Tile size $N_T = 16$ is therefore dominated by the optical power contribution, $P_{\text{neuron}} = P_E + P_\lambda = 2.516$ mW. Assuming a case of continuous 1 GSpike/s spike firing rate, this corresponds to 2.516 pJ/spike. Furthermore, we can consider an illustrative case of a minimal spiking interlink (Figure 8(b)), where the optical spikes from a single neuronal unit (using one MRM) are being optically weighted by a single non-volatile MRM before reaching the photodetectors. If we therefore consider this minimal architecture with one neural MRM and one weighting MRM (with the same IL = 0.2 dB as in all the other cases) and target the same peak optical power per WDM channel $P_\lambda$ at the PD stage which enabled good performance of the full O/E-SNN, we obtain required $P_\lambda \approx -14$ dBm, yielding $P_{\text{neuron}} = P_\lambda + P_E = 44$ uW and 44 fJ/spike at 1GSpike/s.

While our full O/E-SNN architecture also includes weighting blocks, we can make an idealistic assumption of non-volatile tuning of MOSCAP MRMs in the weight bank, with an almost zero static power consumption [45] in an ideal case (without weight bank reprogramming). MOSCAP MRMs have also been shown with sub-volt tuning [46], further strengthening their prospects for optical computing architectures. If we therefore consider an inference-heavy workload with a weight-stationary dataflow (such as the classifier in this work), we can primarily focus on the power consumption from the neural modalities. We also want to emphasize that our O/E-SNN processing architecture is fully analog, and therefore ADC-less, thereby avoiding one of the components incurring significant power consumption in optoelectronic computing [22].

### 3.4. Comparison with related works

We have selected a set of representative published works focusing on optoelectronic and all-optical devices and circuits within the context of SNNs. Table 2 summarizes the comparison in terms of neuron design, possible or proposed network architecture by the reference, neuron dynamics type, maximum firing rate of the neuron, energy per spike, and scalability score of the implementation. Each implementation is graded according to five categories for scalability score: footprint, packaging, WDM, maturity of the fabrication, and cascadability without additional resources (see Supplementary information for more details on the scoring).

Table 2: Comparison between selected neuron designs and network architectures

| HW Type | Design | (Possible) Network | Neuron Type | Max Firing Rate | Energy (pJ/spike) | Scalability Score |
|---|---|---|---|---|---|---|
| Optoelectronics | Memristive junction [47] | digital weighting | LIF-like | 10 KSp/s | 100 | ★ |
| | CMOS IC+VCSEL [48], [49] | MZI Mesh | programmable | 1 GSp/s | 1.18 | ★★ |
| | elect. contr. PCM optical switch [50] | nonl. tuning of add-drop MRM array | thermodynamic LIF-like | 50 MSp/s | 750 | ★★★ |
| | MRM with fb. [51] | MRM weight bank | resonate-and-fire | ~1.1 GSp/s | 10.9 | ★★★★ |
| | p-n MRMs [18] | MRM weight bank | lim. to MRR nonlin. dyn. | 250 MSp/s | 20 | ★★★★★ |
| | **This work** | MRM weight bank | programmable | ~1 GSp/s (28 nm) | 2.5 ($N_T$ 16) 0.044 (min) | ★★★★★ |
| All-optical | inj. locked VCSELs [52] | time delay reservoir | LIF-like | 10 GSp/s | 0.05 | ★★★ |
| | membrane III-V/Si [53] | possibly MRM weight bank | lim. to mode hopping dyn. | 12.5 GSp/s | 1 | ★★★ |
| | two-sect. InP laser [37], [54] | possibly MRM weight bank | limited to laser dynamics | ~1 GSp/s | 50 | ★★★ |
| | Graphene-on-Si MRR [55] | possibly MRM weight bank | integrate-and-fire only exc. | ~40 GSp/s | 0.7 | ★★★ |
| | two-sect. nanolasers [56], [57] | incoherent crossbar array | limited to laser dynamics | ~1 GSp/s | 0.5 | ★★★★ |

In terms of all-optical spiking nodes, spiking can be achieved through various means, either by relying on nonlinear dynamical responses in lasers via injection locking [27], symmetry breaking in resonators [58], or by using multi-section laser devices [37]. All-optical spiking devices offer the highest spiking operation speeds, with the highest firing rates reported from membrane III-V/Si integrated lasers exceeding 12.5 GSpike/s [53]. However, scalability typically represents a challenge. On-chip lasers, as active photonic devices, suffer from complex integration challenges and are often bulky components (up to mm$^2$/device footprint). Injection locking of lasers requires precise control of the input optical signal (wavelength, power, polarization), which hinders scalability, and presents limitations with respect to fan-in from other coherent photonic neurons. Due to vertical light emission, vertical cavity surface emitting laser (VCSEL)-based neurons face packaging challenges. Wavelength control (for WDM) realized with multiple individual lasers adds a further degree of complexity. Furthermore, all-optical spiking methods typically do not offer neural heterogeneity, as well as other modalities such as layer-wide neural inhibition and winner-take-all firing, which can be implemented by electronic spiking circuits. Similarly, the free-space photonic links [47] require bulky setups. Multiport interferometer network architectures are typically aimed at coherent operation, and are less suitable for WDM. At the time of this manuscript's writing, graphene, phase change materials (PCM), and memristive junctions do not rely on mature fabrication platforms. Lastly, even though multiple references reported their neurons' cascadability, the fan-out of a single neuron's output is only reasonable at moderate levels, ~10-15 post-synaptic neurons, due to limited output spike ER, the coupling and propagation losses, excitability threshold of the neurons, etc. Therefore, these architectures require an amplification stage in between the SNN layers. We conclude that only the optoelectronic neurons with closely integrated MRMs are marked in the *'cascadable without additional resources'* category. Furthermore, we envision that monitoring of neural state variables (like membrane potentials) in electronic neurons is likely to be practically simpler than monitoring of carrier dynamics in lasers, which might prove beneficial for experimental SNN training procedures.

In summary, we have introduced `SEPhIA` – an integrated optoelectronic WDM-enabled SNN hardware architecture that combines excitable analog CMOS circuits with compact, integrated photonic devices (MRMs) and multi-wavelength lasers shared per multiple neurons. By considering scalability as a key enabling principle, we have devised the optoelectronic tile (Op-Tile) as a functional building block of our O/E-SNN, and evaluate the scaling of these tiles from the perspectives of both realistic optical power budgets and frequency domain limitations. Using the estimated realistic optical tile size ($N_T = 16$), we have hardware-software co-designed the `SEPhIA` architecture as a multi-tiled, structurally-sparse, feed-forward, all-MRM-based WDM-enabled O/E-SNN architecture. We have validated this architecture both at the device (neuron) level, as well as a two-layer SNN classifier using physics-aware BPTT training. We have demonstrated the operation of the O/E-SNN classifier using a widely-used academic benchmarking task (four-class image classification) with rate-based encoding of feature values. Furthermore, we have demonstrated how photonic device parameters influence the performance (a classification accuracy) at a full SNN level. Finally, we have provided a comparison with other spiking neuromorphic approaches, demonstrating favorable energy/spike metrics. We believe that our O/E-SNN architecture provides a solution to neuromorphic photonic computing that is practically feasible and realistically scalable while offering high computational expressivity for solving non-trivial (temporal) tasks. For future work, we envision a broad set of characteristics that can be further incorporated and explored in the simulations, including quantization (limited bit precision) of photonic modulator states, additional noise sources and different analog neural circuits, among others; as well as more advanced methods for introducing layer sparsity or use of datasets that are natively temporal, spike-based or in the optical domain, such as various time-series data, sensor data or telecommunication data streams.

## 4. Methods

### 4.1. Verilog-A model of the O/E neuronal unit

To facilitate co-simulation of the MRM with electronic components, we have implemented compact Verilog-A models. While numerous approaches have been explored for integrating photonic devices such as MRMs and photodetectors with electronic integrated circuit design, embedding photonic models directly in SPICE (which is widely used for circuit design) offers a more efficient workflow, saving significant time and effort during the design phase [59].

The AdEx LIF model Figure 2(a) was implemented in TSMC 28nm process, and is modeled as a set of coupled differential equations. The membrane potential, $V_{\text{mem}}$, dynamics are governed by a balance between capacitive charging through the membrane capacitance, $C_{\text{mem}}$, leakage currents controlled by the leakage conductance, $g_{\text{leak}}$ along with the resting potential, $E_{\text{L}}$, an adaptation current, $w$ and externally applied input current, $I_{\text{in}}$. The spike is initiated first through an exponential term parameterized by the threshold potential $V_T$ and the slope factor, $\Delta_T$.

$$C_{\text{mem}} \frac{dV_{\text{mem}}}{dt} = I_{\text{in}}(t) - g_{\text{leak}}(V_{\text{mem}} - E_{\text{L}}) + g_{\text{leak}} \Delta_T \exp\left(\frac{V_{\text{mem}} - V_T}{\Delta_T}\right) - w$$

Adaptation is introduced through the current $w$ which evolves with a time constant $\tau_w$. Parameter $a$ controls the strength of the subthreshold depolarization that drives the adaptation, while $b$ determines the discrete increment of adaptation that occurs after each spike.

$$\tau_w \frac{dw}{dt} = a(V_{\text{mem}} - E_{\text{L}}) - w$$

The membrane potential, $V_{\text{mem}}$ is reset to its reset potential $V_{\text{reset}}$ when it crosses its threshold voltage while the spike triggered adaptation is governed by

$$w = w + b$$

As the adaptation grows, it provides a negative feedback to the membrane potential, shaping the temporal structure of spiking activity. The interaction between these elements produces a diverse set of spiking behaviors [60] (Figure 2). This neuronal model is further coupled to the MOSCAP MRM circuit.

The electrical subcircuit of the MOSCAP MRM represents the gated waveguide section as a bias-dependent MOS capacitance, $C_{\text{MRM}}(V)$, derived from accumulation-depletion-inversion behavior, in line with any additional parasitic capacitance. This capacitive load is driven through a series access resistance, $R_{\text{MRM}}$. A dedicated internal node, $V_{\text{dynamic}}$, is introduced between the ExIC and the MOSCAP to account for the RC-limited charging behavior, ensuring that the voltage used for refractive index and absorption change calculations reflects the actual gate dynamics rather than the ideal drive waveform. The bias-dependent capacitance $C_{\text{MRM}}(V)$, effective index change $\Delta n_{\text{eff}}(V)$, and absorption change $\Delta\alpha(V)$ are obtained from polynomial fits to measured [61] or simulated device data, and carrier-density data using plasma dispersion and free-carrier absorption relations. The optical submodel employs time-domain coupled-mode theory to describe the evolution of the intracavity field amplitudes with resonance detuning determined by $\Delta n_{\text{eff}}$ and round-trip loss modified by $\Delta\alpha$. This framework enables accurate prediction of the modulators' through- and drop-port optical powers under arbitrary electrical drive signals, while preserving physical parameters such as ring radius, coupling coefficient, and quality factor of microrings.

Furthermore, our model considers a closed-loop feedback with an additional PD at the drop port of the MRMs (see Figure 2(b)). In WDM systems, where adjacent channel resonances can lead to significant crosstalk, such closed-loop bias stabilization is essential. Our work extends prior demonstrations of automatic wavelength stabilization via bias control of MRMs [62], which dynamically adjusts the ring bias to align and hold the resonance wavelength per channel under varying operating conditions. The high-speed electrical bias tuning in our design enables each ring to be locked to its optimal wavelength with minimal latency, ensuring stable, low-crosstalk performance across all channels. By actively stabilizing the microring resonance through electrical bias modulation and monitoring, our approach avoids the slower response of purely thermal tuning methods while maintaining robust wavelength alignment across all channels.

### 4.2. Training of the O/E-SNN model

As highlighted in Figure 1, the current `SEPhIA` architecture utilizes a standard MRM weight bank design [63] with all-pass MRMs. The odd rows of the MRM weight bank realize excitatory synaptic functionality, while the even rows realize inhibitory synaptic functionality. The trainable parameters of the model directly correspond to the individual resonance wavelength detuning $\Delta\lambda_{\{\tau,\text{r},\text{c}\}}$ of each all-pass $r$-th row, $c$-th column MRM in $\tau$-th Op-Tile. To effectively limit the MRM resonance shifting range, the model utilizes sigmoid-based clamping of the trainable weights to the desired MRM resonance wavelength shifting interval denoted $\Delta\lambda_{\text{max}}$ as $\Delta\lambda_i = \Delta\lambda_{\text{max}} \cdot \sigma(W_i)$. In all cases, positive weight assumes a negative MRM resonance wavelength shift (resonance blue-shift). For both weighting and neural MRMs, we assume an ideal case of low, fixed insertion loss (IL) of 0.20 dB following state-of-the-art for silicon MRMs [64].

In addition to MRMs, a second set of trainable parameters $G$ models a programmable electrical gain within each neuronal unit, and is intended to provide an (optional) trainable degree of freedom in the model for compensation of lack of optical gain blocks in the signal pathway. The ExIC functionality is modelled using a block of Leaky Integrate and Fire (LIF, `snnTorch.Leaky()` [28]) neurons, which were extended to optionally exhibit an *absolute refractory period* to implement some aspects of the adaptive model. The refractory period can be optionally specified in the number of timesteps.

We train the O/E-SNN using a fused `AdamW` optimizer with cosine learning rate scheduling (targeting `lr` $= 0$ at the end of the final epoch). We use the optimizer without weight decay, as we have observed better achieved accuracies without the additional L2 regularization term. We use cross-entropy loss (`CELoss`) with decoding based on max-over-time-membrane potential (MOTM) [28], where the membrane potential of the desired class is encouraged to increase (and vice versa for the non-desired classes). By extension, this maximizes the number of spikes over the simulation run time at the neuron corresponding to the desired class. Alternative coding approaches include end-over-time-membrane (EOTM) or time-to-first-spike (TTFS) [65], among others. Prior to the neuronal units, we utilize conventional `Dropout` blocks ($p = 0.15$) to increase robustness of the neural network training. The explored O/E-SNN has a total of 660 trainable parameters, with 640 parameters corresponding to positive-valued resonance shifts of the programmable MRMs in the weight banks (320 for excitatory and 320 for inhibitory connections).

The limited signal-to-noise ratio present in the model is primarily implemented at the photodetector level, which incorporates two noise effects: the thermal (Nyquist-Johnson) noise with variance $\sigma^2_{\text{thermal}} = 4K_B T \frac{f_{\text{cut}}}{R_{\text{load}}}$ and the shot noise with variance $\sigma^2_{\text{shot}} = 2e f_{\text{cut}}(I_{\text{PD}} + I_{\text{dark}})$. The simultaneous effect of the two noise effects is modelled at every timestep as a Gaussian white noise with variance described as a sum of the individual effect variances. Our simulator currently does not model any thermal crosstalk effects among photonic devices.

# Supplementary information:

# `SEPhIA`: < 1 laser/neuron Spiking Electro-Photonic Integrated Multi-Tiled Architecture for Scalable Optical Neuromorphic Computing

## Appendix A. Description of the O/E-SNN model

### A.1. Compact MRM model in the O/E-SNN

For all the all-pass MRMs in the O/E-SNN models (both the neural MRMs and weighting MRMs), we represent each MRM using its input-output transmission (attenuation) characteristic. This compact device model assumes the MRM through-port transmission as an ideal, Lorentzian-based notch filter. Assuming $f_{\text{reso}}$ is the resonance frequency of the MRM and $Q_{\text{MRR}}$ is the $Q$-factor of the MRM, the notch filter spectral full-width at half-maximum (FWHM) denoted as $\Gamma = \frac{f_{\text{reso}}}{Q_{\text{MRR}}}$.

Assuming $f_{\text{reso}}$ is the set resonance frequency of the MRM, $f$ is the investigated optical frequency (corresponding for example to a given WDM channel), $T_{\text{ER}}$ is the relative through power at the maximum extinction ratio (ER), $T_{\text{IL}}$ is the relative pass-through power accounting for device insertion loss (IL), $\Gamma/2$ is spectral half-width at half-maximum (HWHM), then the attenuation of the MRM at a through port (denoted as $T_{\text{MRM}}$) is modelled:

$$T_{\text{MRM}} = \left(1 - \frac{(\Gamma/2)^2}{(f_{\text{reso}} - f)^2 + (\Gamma/2)^2} \cdot (1 - T_{\text{ER}})\right) \cdot T_{\text{IL}}$$

Since the system is operated in a WDM, non-coherent fashion, phase-shifts at a component level are not considered in the current model. We also currently do not consider the additional passbands (outside of the main resonant frequency) introduced by the MRM's finite free spectral range (FSR) during individual device modelling, but we always ensure that an individual Op-Tile operates within a single FSR range of used MRMs. We assume that all the all-pass MRMs in the whole architecture (that is, both in the neural layer as well as in the weighting layer) have equivalent properties, namely their $Q$-factor, ER, and IL, and also assume a fixed notch filter shape (that is, fixed $Q$-factor and ER) during the MRM tuning. In the presented results, we assume continuously tunable MRM resonances of the MRMs in the weight bank, that is, without resonance detuning (modulation) quantization.

### A.2. O/E neuronal unit

The O/E-SNN consist of neural network layers, and each neural network layer contains one or more Op-Tile(s). From a functional perspective, the nonlinear activations between neural network layers are realized with O/E neuronal units. The forementioned set of microring resonator modulators (MRMs) within a given Op-Tile also forms part of the neuronal units between the given layer and its preceding layer.

An O/E neuronal unit consists, in the respective signal flow order, of a pair of balanced photodetectors (BPDs) for non-coherent O/E conversion of upstream (spiking) signals with excitatory and inhibitory functionality, an (optional) transimpedance amplifier (TIA) or other form of amplifier (gain source), the ExIC spiking circuit, and a (neural) MRM. The balanced photodetector is modelled as a pair of simplified PDs converting light signals represented as complex electric field $E$ to current $I_{\text{PD}} = R_\lambda |E|^2$, where $R_\lambda$ corresponds to the photodetector responsivity. Furthermore, the simplified PD incorporates multiple noise sources, as described in the Methods section.

In the adaptive exponential (AdEx) leaky integrate-and-fire circuit, photodetectors provide excitatory or inhibitory input currents that are integrated onto the membrane capacitance. This capacitance represents the neuronal membrane, where charge accumulation governs the membrane potential, $V_{\text{mem}}$. The membrane node is coupled to an operational transconductance amplifier (OTA), which is biased to act as a tunable conductance, thereby realizing the leak term of the model. The exponential nonlinearity required for spike initiation is achieved by exploiting the subthreshold characteristics of an nFET. Specifically, the membrane voltage, $V_{\text{mem}}$, is compared against a programmable threshold voltage, $V_{\text{T}}$. The OTA amplifies the difference, and the resulting output controls the gate-source voltage of the nFET operating in subthreshold regime. As $V_{\text{mem}}$ approaches $V_{\text{T}}$, the exponential current growth in the nFET drives the rapid upstroke of the spike. Following spike initiation, adaptation dynamics are introduced through an integrator stage. Each time a spike occurs, a fixed amount of charge is injected into the adaptation capacitance, $C_{\text{adapt}}$. This generates a spike-triggered current that feeds back into the membrane node, gradually hyperpolarizing the neuron and reducing its excitability. This adaptive mechanism allows the circuit to reproduce firing behaviors such as adaptation and bursting.

The output of each CMOS neuron is considered as an electrical signal directly driving a single, corresponding integrated neural MRM. Each neural MRM acts upon a single wavelength channel from the comb laser. Therefore, each neural MRM performs simultaneously wavelength DEMUX-ing and on-off keying (OOK) of spikes from its corresponding spiking CMOS neuron. These MRMs therefore operate in a volatile fashion. During the steady (no spike) state, we assume all the MRM resonance wavelengths are perfectly aligned with their corresponding WDM channels from the frequency comb source, effectively minimizing the optical power passing through the neuronal MRMs. A set of parallel neurons in a single layer is realized using the forementioned O/E neuronal units, with all the MRMs sharing a single bus waveguide.

Furthermore, if we assume an add-drop MRM, an additional feedback loop circuit can be implemented for MRM wavelength stabilization via bias control [62]. In our proposed implementation, the drop port of the MRM is coupled to an on-chip monitoring photodetector that continuously generates an electrical signal proportional to the transmitted optical power. This signal is processed by a feedback control loop that adjusts the DC gate bias applied to the MOSCAP-based MRM, compensating for resonance drift caused by temperature fluctuations or slow environmental changes. The control algorithm introduces a small dither around the nominal bias and uses the resulting modulation in the photodetector output to determine both the sign and magnitude of detuning. The feedback loop then drives the bias toward the point of maximum slope, ensuring optimal modulation efficiency and minimizing insertion loss over time. This closed-loop approach enables long-term stability of the MRMs operating point under varying conditions without the need for manual retuning.

### A.2.1. Op-Tiles

As mentioned previously, each Op-Tile contains **(a)** a single, multi-wavelength optical source (such as a comb laser), **(b)** a set of (neural) microring resonator modulators, followed by **(c)** MRM-based, WDM-enabled integrated photonic signal routing and weighting circuit. The multi-wavelength source can either be an on-chip multi-wavelength laser (such as a frequency comb source) or coupled from off-chip, in both cases effectively realizing a neuromorphic photonic system (Op-Tile) with < 1 laser/neuron. Each Op-Tile can have its own light source, or light from a single source can be shared among multiple tiles (power permitting). The single array of neural MRMs was described in the previous section. Following that, an MRM bank with the broadcast-and-weight protocol [26] is utilized to perform amplitude weighting of the WDM-encoded optical spikes. We assume ideal power splitting, that is, for 1:N power splitter, each output of the splitter receives $\frac{1}{N}$ optical power. Here, we want to emphasize that the architecture is not strictly reliant on the weight bank design, and can utilize other photonic architectures for WDM-enabled weighting, such as microring-based crossbars [44]. Furthermore, the use of add-drop MRMs is also a viable alternative option, which practically reduces the number of trainable parameters in the weighting layer by half, but also represents additional constraints due to excitatory-inhibitory weight coupling and imbalancing due to additional losses intrinsically present at the MRM drop port.

## A.3. Model comb-source spectrum

In Figure 3, a numerically generated multi-wavelength laser (frequency comb) spectrum is used. This simplified model implements the multi-wavelength source using analytical description of a uniformly spaced set of Lorentzian-shaped lines. The power levels for all the lines are fixed across all the O/E-SNN model runs (these power levels are read from a pre-computed file that is generated by sampling from uniform distribution to define a flat-top power distribution over a provided dB-band). Full-width half-maximum (FWHM) for all the spectral lines was 8 GHz, and optical powers are sampled from a 2 dB band.

# Appendix B. Additional details on the training

## B.1. O/E-SNN model initialization

We initialize the O/E-SNN by sampling random weight values from uniform distribution $W_{\text{exc}} \sim \text{U}(-2, 2)$ for W corresponding to MRMs in the excitatory branches of the weight bank, $W_{\text{inh}} \sim \text{U}(-3, 1)$ for W corresponding to MRMs in the inhibitory branches of the weight bank, and $G \sim \text{U}(-0.9, 1.1)$. As mentioned in the main text, these weight values $W$ are converted to the MRM resonance shifts $\Delta\lambda_{\text{MRM}}$ as:

$$\Delta\lambda_{\text{MRM}} = \Delta\lambda_{\text{max}} \cdot \sigma(W)$$

where $\Delta\lambda_{\text{max}}$ is the maximum MRM resonance wavelength shift value.

## B.2. Used dataset for the O/E-SNN

To test our proposed O/E-SNN, we utilize a subset of the widely used image classification dataset of fashion articles [34], due to the well-understood nature of the dataset and ease of performing dimensionality reduction on the single-channel (grayscale) image data, allowing us to benchmark our realistically-scaled O/E-SNN model. The used subset includes four classes: `0` (T-shirt/top), `1` (Trouser), `4` (Coat), and `5` (Sandal). We reduce the dimensionality of images from 784 ($28 \times 28 \times 1$, where 1 accounts for single grayscale channel) to 32 using principal component analysis (PCA). Furthermore, we create a rate-encoded representation of these features (where each normalized feature is used as the probability of a spike occurring at any given time step) using `snntorch.spikegen.rate` [28]. A new rate-coded representation is created every time the data is sampled from the dataset. The explained variance of the dataset using principal component analysis (PCA) is 82.61% for $n = 32$ principal components of the Fashion dataset. Alternative approaches to spike-encoded classification benchmarking datasets have also been previously reported, for example in [66]. In summary, we utilize a dimensionally reduced, rate-coded image-based dataset, which can be considered as feeding the O/E-SNN with the outputs of a small-scale neuromorphic vision sensor. We also want to remark that our goal was not to propose our architecture as an optimal solution to the specific classification problem, but rather to demonstrate the operation of our O/E-SNN using a widely understood, non-trivial classification dataset with convenient methods available for dimensionality (feature count) reduction.

## B.3. Comparison to classical SNN models

For the comparison (baseline) to classical SNNs, we utilize two feed-forward SNN models built with `snnTorch`, with no photonic components:

- the FC-SNN (fully connected) has weight matrices of sizes $((36 \times 18), (18 \times 4) = 720$ float32 parameters),
- the sparse SNN (with structured sparsity) incorporates weight matrices consisting of two block-diagonal sub-matrices sized $18 \times 9$ in the first layer, and an $(18 \times 4)$ weight matrix in the second layer, or a total of 396 float32 parameters.

Both of these SNN models have been designed to be in comparable in scale to the O/E-SNN model.

## B.4. Parameters of the O/E-SNN simulation

Table 3: Simulation parameters.

| Category | Parameter | Symbol | Value | Unit |
|---|---|---|---|---|
| O/E-SNN | total trainable parameters | | 660 | - |
| | layer count | $\zeta$ | 2 | - |
| | Op-Tiles count per layer | $\tau$ | (2,1) | - |
| | O/E-SNN timesteps | $T$ | 35 | - |
| Photonics | WDM channel max. possible optical power | $P_\lambda^{\max}$ | 6 | dBm |
| | WDM channel peak optical powers | $P_\lambda$ | $\sim \mathrm{U}(P_\lambda^{\max} - 2, P_\lambda^{\max})^*$ | dBm |
| | WDM channel spacing | $\Delta\omega$ | 100, 63, 50 | GHz |
| | MRM(WB) reso. shift range (at 100GHz) | $\Delta\lambda_{\max}$ | [0,−400] | pm |
| | MRM(N) reso. shift (at 100GHz) | $\Delta\lambda_{\mathrm{optimal}}$ | −335 | pm |
| | MRM(WB) reso. shift range (at 63GHz) | $\Delta\lambda_{\max}$ | [0,−250] | pm |
| | MRM(N) reso. shift (at 63GHz) | $\Delta\lambda_{\mathrm{optimal}}$ | −210 | pm |
| | MRM(WB) reso. shift range (at 50GHz) | $\Delta\lambda_{\max}$ | [0,−200] | pm |
| | MRM(N) reso. shift (at 50GHz) | $\Delta\lambda_{\mathrm{optimal}}$ | −165 | pm |
| | MRM $Q$-factors | $Q$ | 10000, 5000 | - |
| | MRM extinction ratios | ER | 15, 10 | dB |
| | MRM insertion losses | IL | 0.2 | dB |
| | PD responsivity | $R_\lambda$ | 0.5 | A/W |
| | PD frequency cutoff | $f_{\mathrm{cut}}$ | 2.5 | GHz |
| | PD temperature | | 300 | K |
| | PD dark current | $I_{\mathrm{dark}}$ | 1 [67] | nA |
| | PD load resistance | $R_{\mathrm{load}}$ | 50 | Ω |
| LIFs (model) | LIF beta parameter | $\beta$ | 0.99 | - |
| | LIF refractory period | $T_{\mathrm{ref}}$ | 0 | timesteps |
| | LIF firing thresholds in each layer | | 0.5, 0.25 | timesteps |
| Dataset | Dataset | | *fashion-MNIST* | - |
| | Classes | | (0,1,4,5) | - |
| | training set | | 21760 | images |
| | validation sample count | | 2048 | images |
| | test sample count | | 3900 | images |
| | used PCA components | | 32 | - |
| Training | Training batch size | | 128 | samples |
| | Training epochs | | 15 | epochs |
| | Learning rate scheduler | | cosine annealing | - |
| | Initial learning rate | | $4 \times 10^{-2}$ | - |
| | Final epoch learning rate | | 0 | - |
| | Dropout probability | | 0.15 | - |

---

*Sampled from uniform distribution U.

# Appendix C. Calcuation of Scalability scores

To evaluate the scalability scores of other approaches found in the literature, we have devised a scoring mechanism that takes into account five aspects of the architectures: (a) footprint, (b) packaging, (c) support for WDM, (d) maturity of fabrication technology and (e) neural cascadability without requirement of amplification. The ratings on each of these aspects are shown in Table 4 below.

Table 4: Scoring details for overview of comparable approaches.

| HW Type | Design | Footprint | Packaging | WDM | maturity of fab | cascadability w/o amplifier | Scalability Score |
|---|---|---|---|---|---|---|---|
| Optoelectronics | Memristive junction [47] | ✓ | ✗ | ✗ | ✗ | ✗ | ★ |
| | CMOS IC+VCSEL [48], [49] | ✓ | ✗ | ✗ | ✓ | ✗ | ★★ |
| | electrically-controlled PCM optical switch [50] | ✓ | ✓ | ✓ | ✗ | ✗ | ★★★ |
| | MRRM with fb. [51] | ✓ | ✓ | ✓ | ✓ | ✗ | ★★★★ |
| | p-n MRRs [18] | ✓ | ✓ | ✓ | ✓ | ✓ | ★★★★★ |
| | **This work** | ✓ | ✓ | ✓ | ✓ | ✓ | ★★★★★ |
| All-optical | inj. locked VCSELs [52] | ✓ | ✗ | ✗ | ✓ | ✓ | ★★★ |
| | membrane III-V/Si [53] | ✗ | ✓ | ✓ | ✓ | ✗ | ★★★ |
| | two-sect InP laser [37], [54] | ✗ | ✓ | ✓ | ✓ | ✗ | ★★★ |
| | Graphene-on-Si MRR [55] | ✓ | ✓ | ✓ | ✗ | ✗ | ★★★ |
| | two-sect nanolasers [56], [57] | ✓ | ✓ | ✓ | ✓ | ✗ | ★★★★ |